\documentclass[a4paper, 11pt]{article}

\usepackage[english]{babel}

\usepackage[a4paper,top=2cm,bottom=2cm,left=3cm,right=3cm,marginparwidth=1.75cm]{geometry}

\usepackage{amsmath, amssymb, amsfonts, amstext}
\usepackage{graphicx}
\usepackage{natbib}
\usepackage[colorlinks=true, allcolors=blue]{hyperref}
\usepackage{aas_macros}
\usepackage{bm}
\usepackage{xcolor}
\usepackage{subcaption}
\usepackage{graphicx}

\newcommand{\comment}[1]{} 

\title{\textbf{\LARGE \bm{$f(R)$} gravity in the solar system and cosmological scalarons}}
\author{\small
	Debojit Paul$^{1}$\footnote{\href{mailto:debojitpaul645@gmail.com}{debojitpaul645@gmail.com}} \ 
	and Sanjeev Kalita,$^{1}$\footnote{\href{mailto:sanjeev@gauhati.ac.in}{sanjeev@gauhati.ac.in}}
	\\
	\small $^{1}$Department of Physics, Gauhati University,
 Guwahati-781014, Assam, India\\
}

\date{}

\begin{document}

\maketitle

\begin{abstract}
Since last two decades $f(R)$ gravity theory has been extensively used as a serious alternative of general relativity to mimic the effects of dark energy. The theory presents a Yukawa correction to Newtonian gravitational potential, acting as a fifth force of Nature. Generally speaking, this new force is mediated by a scalar field known as scalaron. It affects orbital dynamics of test bodies around a central mass. When the scalaron becomes massive $f(R)$ gravity reduces to Newtonian theory in the weak field limit. In this paper we investigate scalaron mass in the solar system through existing measurements of perihelion shift of planets, Cassini’s measurement of the Parametrized Post Newtonian parameter and measurement of the Brans-Dicke coupling constant. The scalaron mass is constrained in the range ($9.29\times10^{-18}-5.64\times 10^{-16}$) eV. Our results are consistent with existing constraints on the theory arising from the environment of the Galactic Center black hole and binary pulsar systems. Scalarons realized in the solar system are reproduced in the radiation era (($0.88-53.89$) sec) of the universe with a time varying scalaron mass.
\end{abstract}

\section{Introduction}\label{sec1}

The solar system provides us with a laboratory to test theories of gravitation. Einstein’s general relativity (GR) has been tested by several independent observations and space fly-by experiments \citep{2006LRR.....9....3W}. It has been found to be a remarkably accurate theory of gravity in the scale of the solar system. Testability of several general relativistic effects through very compact orbits of stars near the Galactic Center (GC) black hole (Sgr A*) has been investigated by \cite{10.1093/mnras/stab129}. Schwarzschild pericenter shift of the S2 star near the GC black hole and gravitational redshift of its light have been detected by the Very Large Telescope \citep{gravity2018,gravity2020}. However, in past several decades serious alternatives to GR were proposed to address the primordial singularity problem \citep{STAROBINSKY198099} and to replace mysterious dark matter \citep{10.1111/j.1365-2966.2007.11401.x} and dark energy components \citep{doi:10.1142/S0218271802002025,Starobinsky:2007hu} in the standard model of cosmology with new gravitational physics. Within the framework of GR dark matter and dark energy have to be put by hand to account for origin of large scale structures \citep{1982ApJ...263L...1P,Blumenthal:1984bp} and accelerated expansion of the universe \citep{RevModPhys.75.559}. No laboratory experiment has been able to give satisfactory hint of expected particle candidates of dark matter. For example, direct detection experiments such as PICO \citep{PhysRevD.93.061101} and XENON1T \citep{PhysRevLett.121.111302} have reported null results within their sensitivity ranges. Similar results have been reported by the collider based experiments at LHC as well \citep{Abercrombie:2015wmb}. Strong constraints on dark matter annihilation cross section from the  $\gamma$-ray observations of Milky Way dwarf spheroidal satellite galaxies have also been reported \citep{PhysRevLett.115.231301}. The elusiveness of the dark matter candidates is further supported by several other 
 astrophysical constraints \citep{PhysRevD.102.063017,Chan:2022mkd,PhysRevD.105.123006}. 
 
 \par Dark energy is believed to be a cosmological constant – the repulsive energy density in vacuum with a negative pressure which accelerates the cosmic expansion. This acceleration was independently discovered by \cite{Riess_1998} and \cite{Perlmutter_1999} through observations of distant Type Ia supernovae. The cosmological implications of a vacuum energy component was explored by \cite{doi:10.1142/S0218271800000542} and \cite{Carroll:2000fy}. \cite{RevModPhys.75.559} discussed the roles played by cosmological constant and other dark energy candidates in cosmology. But the energy density of vacuum calculated in quantum field theory \citep{RevModPhys.61.1} is larger than the one derived from the observations of accelerated expansion of the universe \citep{Riess_1998,Perlmutter_1999} by a factor of $10^{120}$. This is realized to be unnatural. It represents a deep puzzle in our understanding of gravitation in cosmological setting. There is a plethora of dark energy models which include dynamical scalar fields carrying negative pressure and coupled dark matter – dark energy scenarios to explain the cosmic acceleration. Interested readers may like to follow excellent reviews available in literature (see e.g. \cite{doi:10.1142/S0218271800000542} for early theoretical models; \cite{RevModPhys.75.559} for the cosmological constant in field theory; \cite{doi:10.1142/S021827180600942X} for dynamical dark energy frameworks; \cite{amendola2010dark} for an extensive review of dark sector in cosmology). Here we emphasize on the point that physics of dark energy is not yet known. Due to these reasons, it is believed that dark matter and dark energy are not exotic forms of matter-energy. These are manifestations of modification of GR in the large scale structure of the universe \citep{10.1111/j.1365-2966.2007.11401.x,PhysRevD.87.064002,ODINTSOV2023137988}.

\par One of the extensively studied extensions of GR is $f(R)$ gravity theory. This is geometric modification of Einstein’s gravitational field equations. In these theories the Ricci scalar, R in gravitational Lagrangian is replaced by a general function, $f(R)$. The Einstein-Hilbert action in $f(R)$ theory is written as (in the unit of $c=1$ and in the Jordan frame where matter is coupled only to the spacetime metric $g_{\mu\nu}$)\citep{amendola2010dark}
\begin{equation}
    S_{EH}=\frac{1}{2 \kappa^2}\int d^4 x \sqrt{-g} f(R) + S_{m}(g_{\mu\nu},\Psi_{m})
\end{equation}
Here $\kappa^2= 8\pi G$ and $S_{m}$ is the matter action that depends on the metric $g_{\mu\nu}$ and matter fields $\Psi_m$. $f(R)$ gravity theory is one of the particular cases of general scalar-tensor gravity theories motivated by low energy string theory, having Einstein-Hilbert action in Jordan frame as (in the unit of $c=1=\kappa^2$)\citep{amendola2010dark}

\begin{equation}
    S_{EH}=\int d^4x\sqrt{-g}[\frac{1}{2}f(\phi,R)-\frac{1}{2}Z(\phi)(\delta\phi)^2]+S_m(g_{\mu\nu},\Psi_m)
\end{equation}

Here $\delta\phi$ is covariant derivative of the scalar field $\phi$ which acts as additional gravitational degree of freedom in the theory. For $f(\phi,R)=f(R)$ and $Z(\phi)=0$ the theory reduces to $f(R)$ theory. For $f(\phi,R)=\phi R$ and $Z(\phi)=\omega_{BD}/\phi$ the theory reproduces Brans-Dicke theory with $\omega_{BD}$ being the Dicke parameter (see the discussion below). String motivated dilaton gravity arises for $f(\phi,R)=2e^{-\phi}-2U(\phi)$ \textbf{and} $Z(\phi)=-2e^{-\phi}$ with $\phi$ being the dilaton field and $U$ being the dilaton potential (See \citet{amendola2010dark} for an extensive review on scalar tensor theories).

\par In cosmological setting of the $f(R)$ gravity theory the gravitational field equations modify the Friedmann-Lemaitre evolution, where a curvature fluid appears leading to self accelerated expansion without adding extra negative pressure sources \citep{doi:10.1142/S0218271802002025,NOJIRI2003147,PhysRevD.70.043528}. In the primordial universe this scenario explains occurrence of inflation without existence of extra scalar field \citep{STAROBINSKY198099}. In the Galactic scales $f(R)$ gravity theory is capable of producing flat rotation curves in Low Surface Brightness (LSB) galaxies without incorporating exotic and hitherto unknown dark matter particles \citep{10.1111/j.1365-2966.2007.11401.x}. These theories contain an additional scalar mode of the gravitational force, known as scalaron and is defined \textbf{by} the derivative $\psi=d f(R)/dR$. Vacuum solution of the $f(R)$ gravity field equations shows that the scalaron field alters the Schwarzschild metric – the weak and static gravitational field around spherically symmetric bodies \citep{2018ApJ...855...70K}. The gravitational potential in $f(R)$ theory contains a Yukawa correction term with $e^{-(M_\psi c r/h)}/r$  variation, where $M_\psi$ is mass of the scalar mode, $c$ is the velocity of light in empty space and $h$ is Planck’s constant. This type of correction is usually known as a fifth force of Nature \citep{PhysRevLett.118.211101,2018ApJ...855...70K}. This gets added to the usual Newtonian term with $1/r$  scaling. Testability of the theory through observation of pericenter shift of compact stellar orbits near the GC supermassive black hole has been extensively investigated by considering astrometric capabilities of existing large telescope facilities such as the Keck, the GRAVITY interferometer in VLT and upcoming Extremely Large Telescopes. \citep{Kalita_2020,Kalita_2021,Lalremruati_2022,doi:10.1142/S0218271823500219}. Recently, a Kerr metric has been constructed in $f(R)$ gravity theory \citep{Paul_2024}. The Kerr metric in $f(R)$ gravity has been found to possess appropriate Schwarzschild limit. For infinitely large scalaron mass black hole solutions of $f(R)$ gravity reduce to those in GR and gravitational potential reduces to Newtonian form. By considering observed bright emission ring of the GC black hole shadow \citep{EventHorizonTelescopeCollaboration_2022} and Lense-Thirring precession of compact stellar orbits near the black hole it has been possible to deduce that 
 $f(R)$ gravity behaves like GR for scalarons with mass in the range $(10^{-17} - 10^{-16})$ eV \citep{Paul_2024}. Yukawa type fifth force of nature with scalar mediator was constrained by radar and optical astrometry of Near Earth Object (NEO) asteroids \citep{Tsai_2023}. The method adopted considers effect of mass of the fifth force mediator on orbital precession of the NEO asteroids. The mass range $(10^{-21}-10^{-15})$ eV corresponding to ultralight fuzzy dark matter regime has been constrained. An important outcome of this study is the prospect for investigating outer regions of the solar system for constraining light scalar mass. 

\par Deviation from Newton's inverse square law characterizes new physics, particularly the one involved in unification of gravity theory and quantum mechanics \citep{Lorenzo_Iorio_2007}. It appears in the form of a Yukawa potential, $U=-(GM\alpha/r) e^{-r/\lambda}$ where $\alpha$ is the Yukawa coupling and $\lambda$ is the Yukawa force range. There is a hypothesis that the Yukawa force range is at least of the order of orbital scale of planets in the solar system. It was tested with the help of corrections to the secular rate of the perihelia of Mercury and Earth. Using the EPM2004 Ephemerides of Mercury and Earth the range of Yukawa force was constrained as $\lambda< 0.18 $ au \citep{Lorenzo_Iorio_2007}. In $f(R)$ gravity this corresponds to a scalar mass $M_\psi > 4.58\times 10^{-17}$ eV in the inner solar system orbits.

\par Scalar field in gravitation theory reminds us of an earlier alternative to GR. It is the Brans-Dicke theory of gravity \citep{PhysRev.124.925}. It was formulated to satisfy Mach’s principle of inertia. As discussed above, it is a class of scalar-tensor theories. These theories contain a long range scalar field ($\phi$) in addition to the spacetime metric tensor which carries the gravitational force. Departure of the theory from GR is quantified by the Dicke parameter, $\omega_{BD}$ which appears in correction terms of the gravitational Lagrangian containing the new scalar field. The weak field limit of the theory reduces to that of GR if $\omega_{BD}$ is infinitely large. The scalar field affects general relativistic prediction of Mercury’s perihelion advance, light deflection near the Sun and gravitational time delay of electromagnetic signals near massive bodies in the solar system \citep{PhysRev.124.925,amendola2010dark}. The theory has been constrained through Cassini’s measurement of time delay of electromagnetic signals by putting a lower bound on the Dicke parameter as $\omega_{BD}>40000$ \citep{2006LRR.....9....3W}. After discovery of accelerated expansion of the universe scalar-tensor theories have been invoked to generalize the cosmological constant into a time evolving dark energy component \citep{PhysRevD.60.043501,bartolo2000quintessence}.

\par Several independent investigations near the GC black hole and in cosmological scales have been able to constrain $f(R)$ gravity theory (see for example, constraints arising from cosmology \citep{doi:10.1142/S0218271811019530,Liu:2017xef};
radio quasars \citep{Xu_2018}; cosmic void velocity profiles \citep{PhysRevD.104.023512}; Type Ia supernovae \citep{Hough:2019dzc}; galaxy clusters \citep{PhysRevD.91.103503}; stellar dynamics near GC black hole \citep{Kalita_2020,PhysRevD.104.L101502,doi:10.1142/S0218271823500219}). In this paper we report constraints on the theory by estimating scalaron mass with the help of existing measurements of perihelion shift of planetary orbits, Cassini’s measurement of the Parametrized Post Newtonian (PPN) parameter and measurement of the Brans-Dicke coupling constant. The scalaron mass constrained in this way is compared with the constraints arising from different astrophysical and cosmological environments such as binary pulsar-white dwarf systems, Big Bang Nucleosynthesis (BBN), GC black hole etc. The papaer is organised as follows. In section \ref{sec1a}, we discuss the general relativistic limit of scalaron gravity. In section \ref{sec2}, we present the bounds on mass of scalarons derived from planetary orbits. In section \ref{sec3}, constraint on scalaron mass from PPN parameter and Brans-Dicke theory is presented. Section \ref{sec4} discusses cosmological implications of scalarons realised in the solar system. In section \ref{sec5a} we compare the scalaron mass bound with constraints arising from measurements in several astrophysical and cosmological environments. Results and discussion are presented in section \ref{sec5}. Finally, we conclude in section \ref{sec8}.

\section{General relativistic limit of scalaron gravity}\label{sec1a}

We use perihelion shift of planetary orbits as probe of scalaron mass. This requires a metric structure. In an earlier investigation a full Kerr-scalaron metric has been derived (see equation (22) of \cite{Paul_2024}) which was shown to reduce to the following standard asymptotic form,

\begin{equation}
ds^2=\left(1-\frac{2m}{r}\right)c^2dt^2+\frac{4ma\sin^2\theta}{r}cdt d\phi-\left(1+\frac{2m}{r}\right)dr^2-r^2\left(d\theta^2+\sin^2\theta d\phi^2\right)+ \mathcal{O}\left(\frac{1}{r^2}\right)
\end{equation}

This metric induces Lense-Thirring precession with precession frequency, $\Omega_{LT}=\frac{GJ_{\odot}}{c^2r^3}$, 
where, $J_{\odot}=M_{\odot}R_{\odot}^2 \frac{2\pi}{P_\odot}$, is the angular momentum of the Sun. Substituting the solar parameters (mass, $M_\odot$; radius, $R_\odot$ and equatorial orbital period, $P_\odot$) and considering Mercury’s orbit ($r\approx 52\times 10^6 $ km) we get Lense-Thirring precession frequency of Mercury’s orbit as $\Omega_{LT}\approx 10^{-20}\  s^{-1}$ which is a tiny effect. Therefore, in this work we consider only Schwarzschild limit of the general metric and investigate the scalaron induced correction to Schwarzschild metric in the solar system. The Schwarzschild-scalaron (SchS) metric appears as a solution of vacuum field equation in $f(R)$ theory for spherically symmetric and static spacetime \citep{2018ApJ...855...70K}. It is expressed as

\begin{equation}\label{eq2}
ds^2=\left[1-\frac{2m}{r}\left(1+\frac{1}{3}e^{-M_\psi r}\right)\right]c^2dt^2-\left[1-\frac{2m}{r}\left(1+\frac{1}{3}e^{-M_\psi r}\right)\right]^{-1}dr^2-r^2d\Omega ,
\end{equation}

 For $M_\psi \rightarrow \infty$ (Compton wavelength of scalaron much smaller than scales of the system concerned, see below), the metric naturally reduces to general relativistic Schwarzschild limit.
It is seen that the gravitational potential in $f(R)$ gravity theory contains a Yukawa correction to Newtonian potential ($m/r$) which has the form, \citep{2018ApJ...855...70K}

\begin{equation}\label{eq1}
\Phi(r)=-\frac{GM}{\psi_o r}(1+\frac{1}{3}e^{-M_\psi r})
\end{equation}

Here, the scalaron mass $M_\psi$ is written in the unit of $c=h=1$. $\psi_o$ is a dimensionless scalar field amplitude in the theory. Since $f(R) = R$ in GR, $\psi_o = df(R)/dR=1$. Also, the mass parameter of the spacetime metric is given by $m=GM/c^2\psi_o$ where $M$ is the mass of the central body.
The gravitational potential can be written in terms of an effective gravitational constant ($G_{eff}$) \citep{amendola2010dark},

\begin{equation}\label{eq6a}
    \Phi(r)=-\frac{G_{eff}M}{r}
\end{equation}

where, 
\begin{equation}\label{eq7a}
    G_{eff}=\frac{G}{\psi_o}\left(1+\frac{1}{3}e^{-M_\psi r}\right)
\end{equation}

It gives the Poisson equation for a mass distribution ($\rho$) as,

\begin{equation}\label{eq8a}
    \nabla^2\Phi=-4\pi\rho G_{eff} 
\end{equation}

In GR $\psi_o=1$. Therefore, Poisson equation reduces to that in ordinary GR if $M_\psi r>>1$. This happens for scales of the system which are bigger than Compton wavelength of scalaron. In linear cosmological perturbation theory of $f(R)$ gravity we have the Poisson equation for sub-horizon scale ($\lambda a<<cH^{-1}$; where $\lambda$ is a co-moving length, $a$ is the scale factor and $H$ is the Hubble parameter) as

\begin{equation}
    \nabla^2\Phi=-4\pi \bar{\rho} G_{eff}^{'} \delta_m
\end{equation}

Here, the effective gravitational constant is defined as $G_{eff}^{'}$ to distinguish it from the one we wrote for static field limit (see equations (\ref{eq6a})-(\ref{eq8a})).  It is expressed as \citep{ShaunA.Thomas_2011}

\begin{equation}\label{eq10a}
    G_{eff}^{'}=G\left(\frac{3+2\frac{k^2}{a^2M_\psi^2}}{3+3\frac{k^2}{a^2M_\psi^2}}\right)
\end{equation}

Also, $\bar{\rho}$ is the background cosmic mass density and $\delta_m=\delta\rho/\bar{\rho}$ is the amplitude of matter density perturbation. The Laplacian operator becomes $\nabla^2=k^2/a^2$ giving rise to Poisson equation as (see for example \citet{Pilar_Ruiz_Lapuente})

\begin{equation}
    k^2\Phi=-4\pi \bar{\rho} G_{eff}^{'} a^2 \delta_m
\end{equation}

From equation (\ref{eq10a}) we observe that if $M_\psi>>k/a$, $G_{eff}^{'}\approx G$ leading to general relativistic (Newtonian) Poisson equation,

\begin{equation}
    k^2\Phi=-4\pi G \bar{\rho} a^2 \delta_m 
\end{equation}

As $k=1/\lambda$ the above condition reduces to $M_\psi r>>1$ with $r=\lambda a$ being the proper length scale of a structure. This is the same condition we encounter while realising GR limit of scalaron gravity in the solar system. The scalaron masses which satisfy this criterion in the solar system are displayed in section \ref{sec5}.

\section{Bound on mass of scalarons from planetary orbits}\label{sec2}

The metric displayed in (\ref{eq2}) has been extensively studied near the GC black hole through its effect on in-plane pericenter shift of stellar orbits \citep{Kalita_2020,doi:10.1142/S0218271823500219} and the black hole shadow measurements \citep{Paul_2024}. Recently, \citet{Paul_2024} constructed a Kerr metric with scalarons and found that it reduces to the SchS metric for zero angular momentum. To constrain scalarons through their effect on orbits of the solar system we follow the differential equation of orbit of a test particle derived earlier for the SchS metric \citep{Paul_2024}. The orbit equation has the following form.

\begin{equation}\label{eq3}
\frac{d^2 u}{d \phi^2}+u=\frac{mc^2}{L^2}\left(1+\frac{1}{3} e^{-\frac{M_\psi}{u}}\right)+3m\left(1+\frac{1}{3} e^{-\frac{M_\psi}{u}}\right)u^2 +\frac{mc^2M_\psi}{3L^2}u^{-1}e^{-\frac{M_\psi}{u}}+\frac{mM_\psi}{3}ue^{-\frac{M_\psi}{u}}
\end{equation}

Here, $m=GM_\odot /c^2\psi_o$ (where $M_\odot$ is the mass of the Sun), $u=1/r$, and $L^2=m a(1-e^2)$ ($a$ being the semi-major axis and $e$ being the eccentricity of the orbits). The first term on the right hand side represents the Newtonian contribution along with its scalaron counterpart ($1/3\ e^{-M_\psi r}$). The second term represents Schwarzschild contribution along with the scalaron correction to the Schwarzschild part. Finally, the third and the fourth term are known as scale modulated scalaron correction terms that arise due to presence of scalarons (see section 3.4 in \citet{Paul_2024} for detailed discussion on these terms). In SchS geometry, the orbit of any test particle undergoes perihelion shift by an amount \citep{Paul_2024}.

\begin{multline}\label{eq4}
(\delta\phi)_{SchS}=(\delta\phi)_{Sch} + \frac{6\pi m}{3a(1-e^2)}e^{-M_\psi a(1-e^2)} + \frac{4\pi m M_\psi}{3}e^{-M_\psi a(1-e^2)}\\
+\frac{2\pi a^2(1-e^2) M_\psi^2}{6}e^{-M_\psi a(1-e^2)}+\frac{2\pi m a(1-e^2) M_\psi^2}{6}e^{-M_\psi a(1-e^2)}
\end{multline}

Here, $(\delta \phi)_{Sch}$ is the Schwarzschild perihelion shift, $6\pi m/ a(1-e^2)$. The appearance of terms proportional to square of semi-major axis is inevitable in presence of Yukawa type fifth force correction to Newtonian gravitational potential \citep{Tsai:2023zza}. The form (\ref{eq4}) is the exact analytical expression for perihelion shift in $f(R)$ gravity theory. The perihelion shift in the above equation is expressed in angle per period. Equation (\ref{eq4}) can be expressed in the following transcendental form.

\begin{equation}\label{eq5}
\left[A M_\psi^2 +B M_\psi + C\right]e^{-M_\psi a(1-e^2)}+D=0
\end{equation}

where,
\begin{equation}
\begin{aligned}
& A =\frac{2\pi a^2 (1-e^2)}{6} + \frac{2\pi m a (1-e^2)}{6}\\
& B = \frac{4\pi m}{3}\\
& C = \frac{6\pi m}{3 a(1-e^2)}\\
& D= (\delta \phi)_{Sch}-(\delta \phi)_{SchS}
\end{aligned}
\end{equation}

\begin{table}
\centering
\caption{Planetary data adopted from literature (The data is adopted from \citet{10.1093/mnras/stac3509} and the references therein; \citet{10.1111/j.1365-2966.2009.16196.x} for semi-major axis, eccentricity, period and \citet{10.1111/j.1365-2966.2009.16196.x} and  \citet{PhysRevD.95.024017} for perihelion shift).}
\label{tab1}
\resizebox{0.8\linewidth}{!}{%
\begin{tabular}{ccccc}
\hline
Planet &
  \begin{tabular}[c]{@{}c@{}}Semi-major axis\\ $a$ (au)\end{tabular} &
  \begin{tabular}[c]{@{}c@{}}Eccentricity\\ $e$\end{tabular} &
  \begin{tabular}[c]{@{}c@{}}Period\\ P (days)\end{tabular} &
  \begin{tabular}[c]{@{}c@{}}Observed perihelion shift\\ $\delta\phi$ (arcsec/century)\end{tabular} \\ \hline
Mercury & $0.3871$ & $0.206$ & $88.97$    & $42.9799_{-0.0006}^{+0.0030}$ \\
Venus   & $0.7233$ & $0.007$ & $224.70$   & $8\pm5$                       \\
Earth   & $1.0000$ & $0.017$ & $365.26$   & $5\pm1$                       \\
Mars    & $1.5237$ & $0.093$ & $686.98$   & $1.3624\pm0.0005$             \\
Jupiter & $5.2034$ & $0.048$ & $4332.59$  & $0.070\pm0.004$               \\
Saturn  & $9.5371$ & $0.056$ & $10759.22$ & $0.014\pm0.002$               \\ \hline
\end{tabular}%
}
\end{table}

The observational bounds on perihelion shift available for planetary orbits up to Saturn\footnote{For Uranus and Neptune reliable data are not available in the literature.} are presented in Table \ref{tab1}. Data from Cassini \citep{bertotti2003test} and Messenger spacecraft \citep{Fienga:2011qh} has given stringent bound on the perihelion shift of Mercury \citep{PhysRevD.95.024017}. This data is combined with data for rest of the five planets \citep{10.1111/j.1365-2966.2009.16196.x,PhysRevD.95.024017}. The observed perihelion shift presented in the Table \ref{tab1} are residual perihelion shift after accounting for all other planetary perturbations, quadrupole moment of the Sun and Lense-Thirring effects \citep{Park_2017}. Therefore, the residual perihelion shift is expected to put constraint on modified gravity effect. On the basis of precession of solar system Keplerian orbits, \citet{PhysRevD.97.104067} produced bounds on strength of the Yukawa correction as $\delta\approx 10^{-2}-10^{-9}$ (for the six planets up to Saturn) where $\delta$ appears in the gravitational potential as,\citep{PhysRevD.97.104067}

\begin{equation}\label{eq1a}
    \Phi(r)=-\frac{GM}{(1+\delta)r}\left[1+\delta e^{-r/\lambda}\right]
\end{equation}

It is to be noted that scalaron mass, $M_\psi$ actually represents mass scale and can be mapped to the parameter $\lambda$ (a length scale) in equation (\ref{eq1a}) (it is to be noted that $1$ au$^{-1}=8.25\times10^{-18}$ eV; See \citet{Kalita_2020} for details on mass scales). The parameter $\delta$ is related to the scalar field amplitude $\psi_o$ in equation (\ref{eq1}) as

\begin{equation}
    \delta=\psi_o-1
\end{equation}

Therefore, the bounds on $\delta$ obtained earlier (see Table I of \citet{PhysRevD.97.104067}) are mapped into bound on $\psi_o$. All the values converge to $\psi_o \approx 1$. Additionally, \cite{PhysRevLett.118.211101} discussed that compatibility of general relativistic pericenter shift of compact stellar orbits near the GC black hole requires coupling constant $\alpha$ (which is related to $\psi_o$ as $\alpha=1/3\psi_o$ \citep{Kalita_2020}) to be $\alpha<0.016$ ($95\%$ confidence limit). We try to investigate the consistency of this result in the solar system as well. Thus, $\alpha=0.016$ corresponds to $\psi_o \approx 20.833$. Hence, the perihelion shift values in Table \ref{tab1} are substituted for $(\delta \phi)_{SchS}$ in equation (\ref{eq4}) and the mass of scalaron is estimated for the two choices of $\psi_o$ ($\psi_o=1$ and  $20.833$).

\begin{table}
\centering
\caption{Constraints on $M_\psi$ in different planetary orbits.}
\label{tab2}
\resizebox{0.7\linewidth}{!}{%
\begin{tabular}{ccc}
\hline
Planet  & \begin{tabular}[c]{@{}c@{}}Scalaron Mass, $M_\psi$ (eV)\\ ($\psi_o=1$)\end{tabular} & \begin{tabular}[c]{@{}c@{}}Scalaron Mass, $M_\psi$ (eV)\\ ($\psi_o=20.833$)\end{tabular} \\ \hline
Mercury & $5.64\times10^{-16}$                                                                & $1.46\times10^{-21}-4.60\times10^{-16}$                                                  \\
Venus   & $2.37\times 10^{-21} - 3.27\times 10^{-16}$                                       & $6.79\times 10^{-21} - 2.38\times 10^{-16}$                                             \\
Earth   & $1.67\times10^{-21} - 2.07\times10^{-16}$                                           & $3.44\times10^{-21}-1.79\times10^{-16}$                                                  \\
Mars    & $1.48\times10^{-16}$                                                                & $1.81\times10^{-21}-1.21\times10^{-16}$                                                  \\
Jupiter & $4.02\times10^{-17}-4.21\times10^{-17}$                                             & $2.93\times10^{-22}-3.73\times10^{-17}$                                                  \\
Saturn  & $2.26\times10^{-17}-2.71\times10^{-17}$                                             & $1.06\times10^{-22}-2.12\times10^{-17}$                                                  \\ \hline
\end{tabular}%
}
\end{table}

Table \ref{tab2} presents the bounds on mass of scalaron obtained for planetary orbits as the solution of equation (\ref{eq5}). It is seen that for the two choices of $\psi_o$ a wide bound on the mass of scalarons is obtained (of the order of ($10^{-22}-10^{-16}$) eV). This mass bound is further examined in the next section from the consideration of scalar-tensor gravity theory.

\section{Constraint on scalaron mass from PPN parameter and Brans-Dicke theory}\label{sec3}

The PPN parameter $\gamma$ measures curvature per unit mass for a massive body (in this case the Sun).  It parametrizes deviation from GR in weak field limit and is measured by experiments on light deflection and Shapiro time delay \citep{2006LRR.....9....3W}. In case of GR, the value of $\gamma$ is unity. In other theories of gravity it deviates from unity \citep{misner1973gravitation}.The PPN parameter in $f(R)$ gravity theory is given by \citep{amendola2010dark,2018ApJ...855...70K},

\begin{equation}\label{eq6}
\gamma=\frac{3-e^{-M_\psi r}}{3+e^{-M_\psi r}}
\end{equation}

In the solar system, $\gamma$ is constrained as $1+(2.1\pm2.3)\times 10^{-5}$. This bound has been given by the measurement of Shapiro time delay performed by the Cassini spacecraft \citep{bertotti2003test}. Using scalaron masses from Table \ref{tab2} and taking perihelion distance of the planets, $r_p=a(1-e)$, the $\gamma$ values are estimated for each of the planetary orbits. The deviations of these estimated values from the observed values ($|\delta\gamma/\gamma|$) have been presented in Table \ref{tab3}. It is seen that for $\psi_o=1$, except for Venus and Earth the deviations of the estimated PPN parameter from the observed bound is extremely small. For $\psi_o=20.833$, all the six planets show deviations from the observed bounds on $\gamma$ for lighter side of the scalaron mass ($\sim 10^{-22}$ eV$-10^{-18}$ eV). Overall, it is observed that the heavier scalarons ($\sim 10^{-16}$ eV$-10^{-17}$ eV) are consistent with the observed bounds on $\gamma$. Now, we narrow down the scalaron mass range from the equivalence of $f(R)$ gravity theory and Brans-Dicke theory (BDT).

\begin{table}
\centering
\caption{The deviation of estimated $\gamma$ from Cassini bounds. The range of deviation (third and fourth column) arises from the scalaron mass range obtained in Table \ref{tab2}.}
\label{tab3}
\resizebox{0.6\linewidth}{!}{%
\begin{tabular}{cccc}
\hline
Planets & $r_p$ (au) & $|\frac{\delta \gamma}{\gamma}|_{\psi_o=1}$ & $|\frac{\delta \gamma}{\gamma}|_{\psi_o=20.833}$ \\ \hline
Mercury & $0.307$    & $2.1 \times 10^{-5}$                        & $0.49 - 2.1\times 10^{-5}$                       \\
Venus   & $0.718$    & $0.49 - 2.09\times 10^{-5}$                 & $0.49 - 2.1\times 10^{-5}$                       \\
Earth   & $0.983$    & $0.49 - 2.09\times 10^{-5}$                 & $0.49 - 2.09\times 10^{-5}$                      \\
Mars    & $1.381$    & $2.09 \times 10^{-5}$                       & $0.49 - 2.1\times 10^{-5}$                       \\
Jupiter & $4.953$    & $2.09 \times 10^{-5}$                       & $0.49 - 2.09\times 10^{-5}$                      \\
Saturn  & $9.003$    & $2.09 \times 10^{-5}$                       & $0.49 - 2.09\times 10^{-5}$                      \\ \hline
\end{tabular}%
}
\end{table}

\par As mentioned in the section \ref{sec1}, Brans-Dicke theory (BDT) is a class of scalar-tensor theories that was studied as the first serious alternative to GR. The theory is of particular interest as in the weak field limit it possesses a structure similar to that of $f(R)$ theory. This occurs due to chameleon mechanism \citep{PhysRevD.69.044026}- a mechanism which allows scalaron mass to be environment dependent so that in the high density regions of planets and the Sun ($\rho \sim 1-10$ g/cc which is larger than the mean cosmological density, $10^{-30}$ g/cc) scalarons become extremely heavy and the gravitational potential (see equation (\ref{eq1})) looks similar to the Newtonian. A detailed theoretical account of environment dependent scalaron mass is presented in section \ref{sec4}. In BDT the gravitational constant, $G$ is expressed in terms of reciprocal of a scalar field $\phi$. The dimensionless Dicke parameter $\omega_{BD}$ measures deviation from GR. The PPN parameter in BDT is given by \citep{amendola2010dark},

\begin{equation}\label{eq8}
\gamma=\frac{1+\omega_{BD}}{2+\omega_{BD}}
\end{equation}

For $\omega_{BD} \rightarrow \infty$, BDT reduces to GR ($\gamma =1$). A very small value of $\omega_{BD}$ ($\sim 0$) represents drastic deviation from GR with $\gamma =1/2$ . This is equivalent to zero scalaron mass in $f(R)$ theory (see equation (\ref{eq6})). From Cassini's measurements $\omega_{BD}$ has been constrained in the solar system as $\omega_{BD} > 40000$ \citep{2006LRR.....9....3W}. Comparing equations (\ref{eq6}) and (\ref{eq8}) the Dicke parameter is expressed in terms of scalaron mass as

\begin{equation}\label{eq9}
\omega_{BD}=\frac{3(1-e^{-M_\psi r})}{2 e^{-M_\psi r}}
\end{equation}

\begin{figure}[hptb]
    \centering
    \begin{subfigure}[b]{\textwidth}
        \centering
        \includegraphics[width=\textwidth]{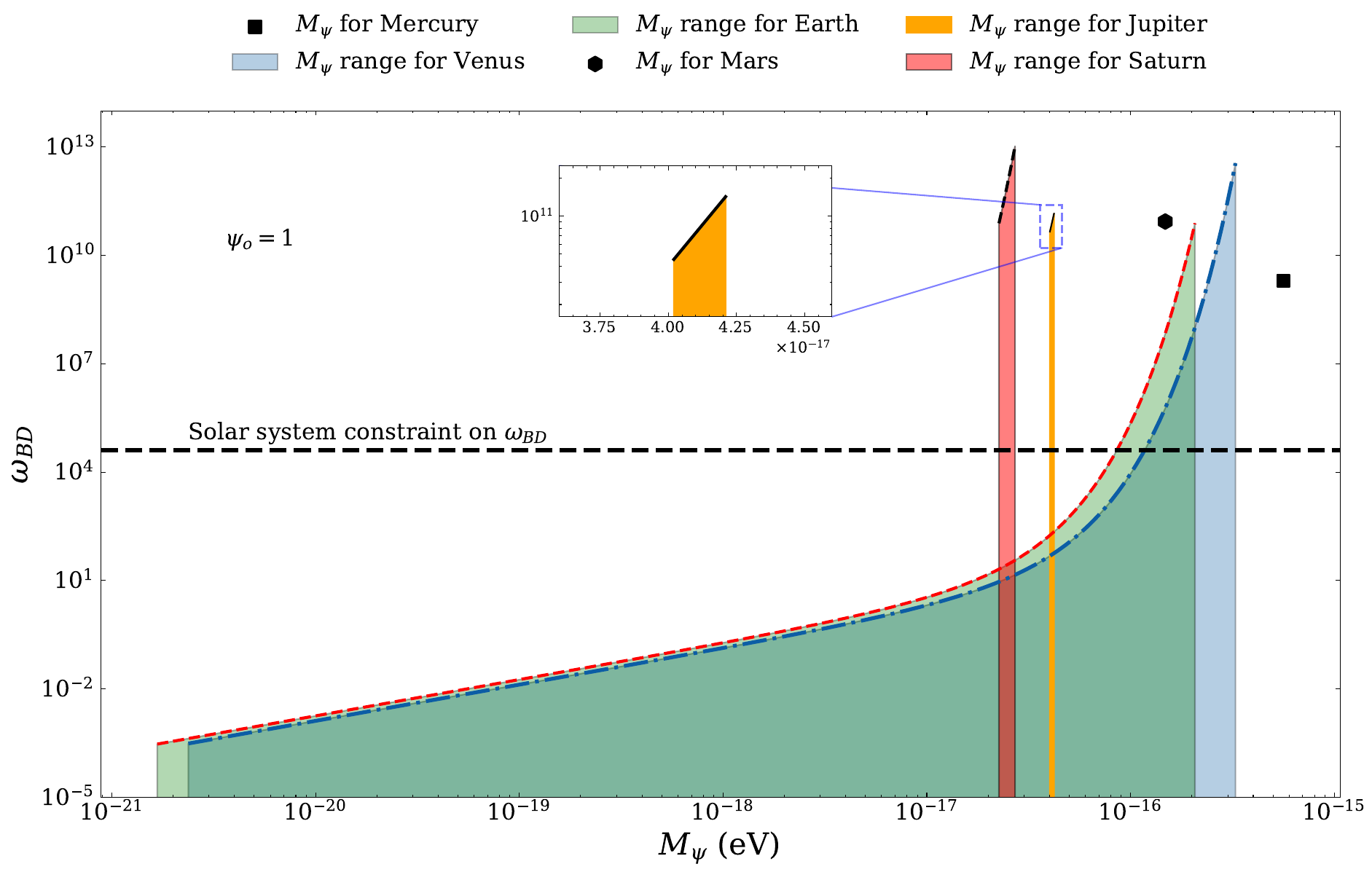}
        \caption{}
        \label{fig:1}
    \end{subfigure}
    \hfill
    \begin{subfigure}[b]{\textwidth}
        \centering
        \includegraphics[width=\textwidth]{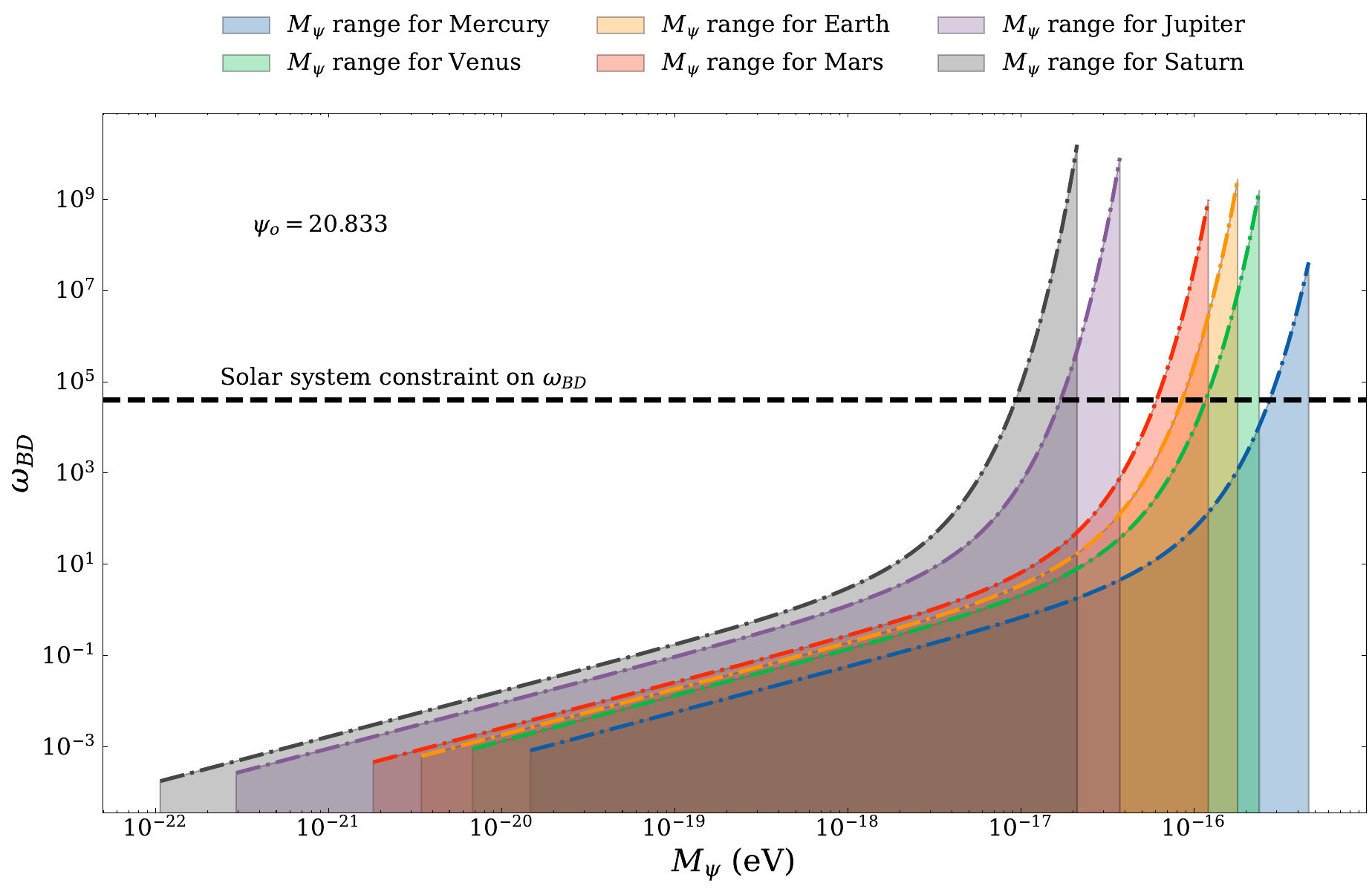}
        \caption{}
        \label{fig:2}
    \end{subfigure}
\caption{Variation of Dicke parameter $\omega_{BD}$ against scalaron mass $M_\psi$ for different planetary orbits for (a) $\psi_o=1$ and (b) $\psi_o=20.833$.}
\label{fig1}
\end{figure}

For perihelion distances of the six planets the variation of $\omega_{BD}$ is studied for different scalaron masses and for two choices of $\psi_o=1\ \& \ 20.833$. The variations are shown in Figure \ref{fig1}. It is seen that for $\psi_o=1$, the scalaron masses derived in Table \ref{tab2} for the orbits of Mercury, Mars, Jupiter and Saturn are consistent with the constraint $\omega_{BD} > 40000$. For the orbits of Venus and Earth much more stringent bounds on $M_\psi$ are obtained. The bound on $M_\psi$ for Venus is ($1.17\times 10^{-16}-3.27\times10^{-16}$) eV and for Earth, the bound is ($8.56\times10^{-17}-2.07\times10^{-16}$) eV. For $\psi_o=20.833$ the $M_\psi$ bound for Mercury is obtained as ($2.75\times 10^{-16}-4.60\times10^{-16}$) eV. For Venus the bound is in the range ($1.16\times 10^{-16}-2.38\times10^{-16}$) eV. For Earth the bound is in the range ($8.47\times 10^{-17}-1.79\times10^{-16}$) eV. For Mars and Jupiter the bounds are in the ranges ($6.14\times 10^{-17}-1.21\times10^{-16}$) eV and ($1.66\times 10^{-17}-3.73\times10^{-17}$) eV respectively. And finally, for Saturn the range is ($9.29\times 10^{-18}-2.12\times10^{-17}$) eV.

\section{Scalarons from cosmology}\label{sec4}

\citet{Yadav_2019} showed that scalarons of power law gravity theory ($f(R)\sim R^m$) have environment (density) dependent mass. $10^{-16}$ eV scalarons were found to be compatible with the density environment of the solar system. These scalarons were interpreted as dark matter. Scalaron of Hu-Sawicki gravity with mass falling with cosmic expansion has been reported in \citet{doi:10.1142/S0217732321502655}. These authors predicted existence of $10^{-23}$ eV scalarons near present epoch. This mass increases towards past of the cosmic history. Scalarons were connected to astronomical phenomena near black holes for the first time in \citet{Kalita_2020}. It was shown that scalaron mass naturally appears from UV and IR scales of curvature induced vacuum fluctuations near black hole horizon. These scales depend on black hole mass. \cite{Talukdar_2024} expressed an explicit relation between scalaron mass and black hole mass. It is a reciprocal relation given by

\begin{equation}\label{eq4a1}
    M_\psi=10^{-10}\ \text{eV}\ \frac{M_\odot}{M}
\end{equation}

A cosmology based on gravitational radius of the universe has been proposed by \citet{10.1111/j.1365-2966.2007.12499.x}. It has been found to be useful for explaining several cosmological puzzles such as time compression problem and appearance of massive black holes at high redshift \citep{melia2018,doi:10.1098/rspa.2015.0449}. The pivotal point of this cosmology is an epoch independent relation between Hubble length ($cH^{-1}$ ) and gravitational radius ($2GM_{Univ}/c^2$) of the universe. They are of the same order. It is easier to visualize this condition from the fact that the universe can itself act as an expanding black hole provided we assume that its Hubble length is equal to or less than gravitational radius. If one takes the Planck epoch as the beginning of the universe it acted as a classical black hole (gravitational radius being equal to or larger than Planck length; see \citet{10.1093/mnras/152.1.75}) of Planck mass, $M_{pl}\sim 10^{-5}$ g. The Hubble length at Planck epoch is the Planck length $c H_{Pl}^{-1} \sim c t_{Pl} \sim 10^{-33}$ cm. This is also the gravitational length of a Planck mass black hole. Therefore, an epoch independent relation $2GM_{Univ} /c^2 = cH^{-1}$ can be adopted. Taking the universe as a black hole and considering the gravitational radius of the Sun ($2GM_\odot/c^2 \sim 3$ km) the scalaron mass can be expressed in terms of the Hubble parameter (expressed in sec$^{-1}$ ) as

\begin{equation}
    M_\psi=10^{-15} \text{eV} \ H \ \text{(sec)}
\end{equation}

We call these scalarons having mass evolving in cosmic time as cosmological scalarons. The linear relationship ($M_\psi \propto H$) between scalar mass and cosmic expansion rate ($H$) is generic in supergravity theory involving light scalar particles. These theories are often invoked for understanding origin of cosmic acceleration. In a class called extended supergravity theory square of scalar mass is proportional to the cosmological constant ($M_\psi^2 \propto \Lambda$). For de Sitter expansion ($\Lambda \propto H^2$) it calls for the relationship $M_\psi \propto H$. These theories can take into account cosmic acceleration with a scalar mass of $10^{-33}$ eV (see \citet{PhysRevD.65.105016,PhysRevD.66.123503} for these considerations). In a simple spatially flat Friedmann like universe the Hubble parameter varies with background cosmological mass density ($\bar{\rho}$) as

\begin{equation}
    H=\sqrt{\frac{8\pi G \bar{\rho}}{3}}
\end{equation}

This gives us density dependent cosmological scalaron mass

\begin{equation}\label{eq18}
    M_\psi=10^{-15} \text{eV} \sqrt{\frac{8\pi G \bar{\rho}}{3}} \ \text{(sec)}
\end{equation}

\begin{figure}
    \centering
    \includegraphics[width=\linewidth]{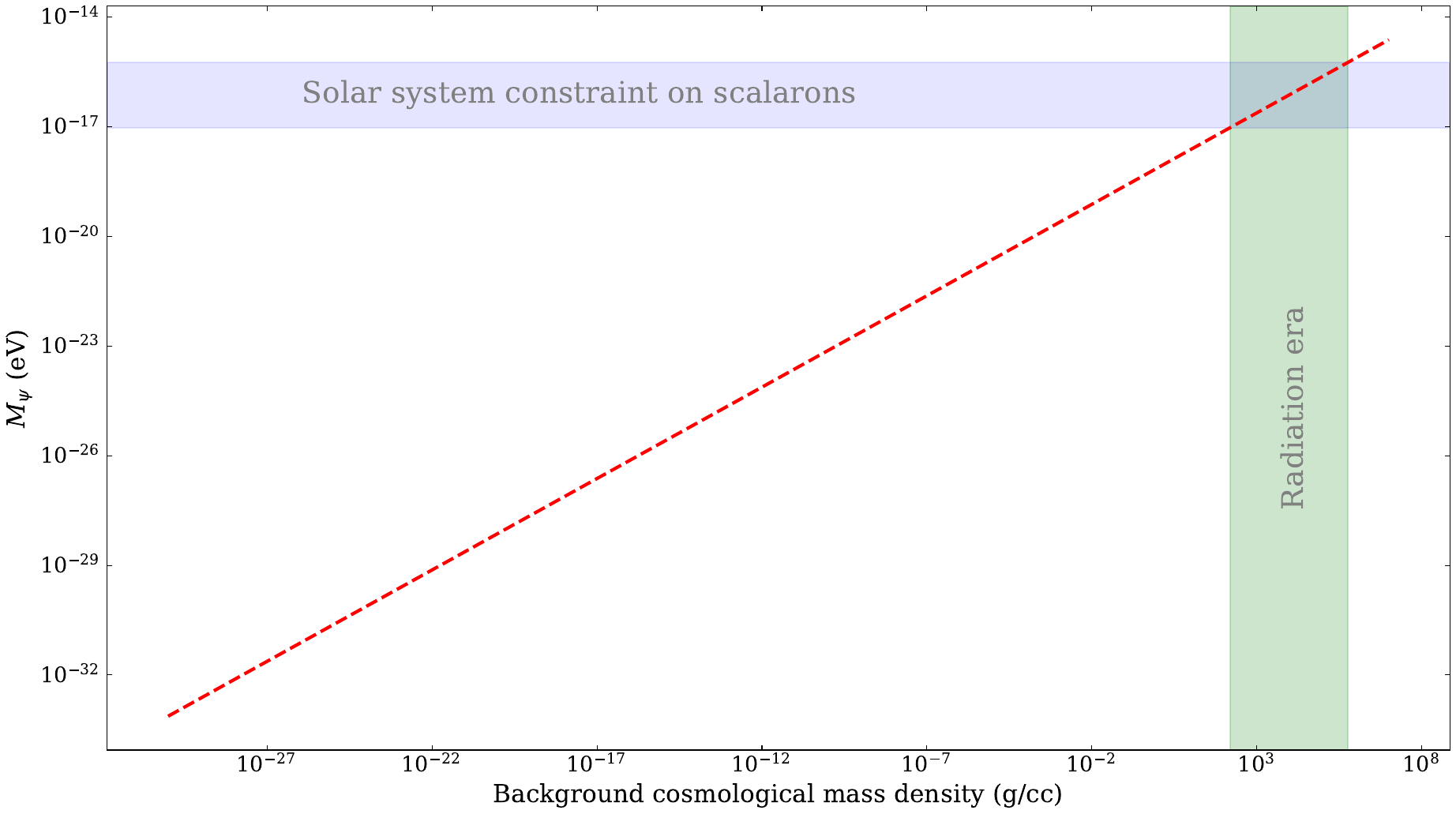}
    \caption{Variation of scalaron mass with background cosmological mass density. The scalaron mass bound (constrained in solar system) ($9.29\times 10^{-18}-5.64\times 10^{-16}$) eV corresponds to density $(1.54\times 10^2 - 5.69\times 10^{5})$ g/cc.}
    \label{fig5}
\end{figure}

Scalaron mass becomes higher in the early universe ($\bar{\rho}$ very large). This relation is based on a black hole based cosmology and is independent of earlier results reporting time variation of scalaron mass. Background mass density of the universe at present epoch is nearly $10^{-30}$ g/cc. This gives a scalaron mass $M_\psi\approx10^{-34}$ eV which is compatible with scalar mass responsible for de Sitter expansion in supergravity theories. Figure \ref{fig5} shows variation of scalaron mass with background cosmological mass density. 

\par The above relation allows us to calculate environment dependent scalaron mass. Considering density perturbation ($\delta\bar{\rho}=\rho-\bar{\rho}$) in the cosmological background, we obtain an environment dependent scalaron mass from equation (\ref{eq18}) as follows,

\begin{equation}  M_{\psi(\text{environ})}=\bar{M_\psi}\frac{\rho}{\bar{\rho}}
\end{equation}

Here, $\bar{M_\psi}$ is the background scalaron mass which is $10^{-34}$ eV, $\rho$ is the environmental density and $\bar{\rho}$ is the cosmological background density. We consider density environment of the stellar interior, for example, center of the Sun where $\rho\approx100$ g/cc. Assuming the Sun to be situated in the cosmological background with $\bar{\rho}\approx 10^{-30}$ g/cc we get

\begin{equation}
    M_{\psi(\text{Sun})}=0.1\ \text{meV}
\end{equation}

In the density environment of the solar system (total planetary mass distributed within a scale of the order of Neptune's orbital size) where $\rho\approx 10^{-12}$ g/cc, scalarons of $10^{-16}$ eV naturally appear. Considering orbital scale of outer Kuiper belt objects ($\sim 200$ au) we reach a density scale of around $10^{-14}$ g/cc. This produces scalarons of $10^{-18}$ eV. These are perturbed scalarons in the solar system scales which have been generated in this study through orbital perihelion shift measurements and other solar system experiments.

\par It is seen that scalarons in the mass range ($9.29\times 10^{-18} - 5.64\times 10^{-16}$)  eV which appears through solar system constraint is reproduced as background scalarons ($\bar{M_\psi}$) in the background density environment $\bar{\rho} \approx (1.54\times 10^2 - 5.69\times 10^{5})$ g/cc (see equation (\ref{eq18})). In the radiation era of the universe the background cosmic mass density evolves with time as,

\begin{equation}
\bar{\rho}=\frac{3}{32\pi G t^2}    
\end{equation}

The above background density environment, therefore corresponds to a slice of cosmic time $t \approx (0.88-53.89)$ sec after the Big Bang. Perturbed scalarons compatible with solar system environment are realized as background scalarons in the radiation era of the universe. These background scalarons have evolved with cosmic time to  today's background scalarons of $10^{-34}$ eV. This is clear indication of time evolution of scalaron masses.

\section{Comparison of scalaron mass with existing astrophysical and cosmological constraints on scalar field mass}\label{sec5a}
\begin{figure}
    \centering
    \includegraphics[width=\linewidth]{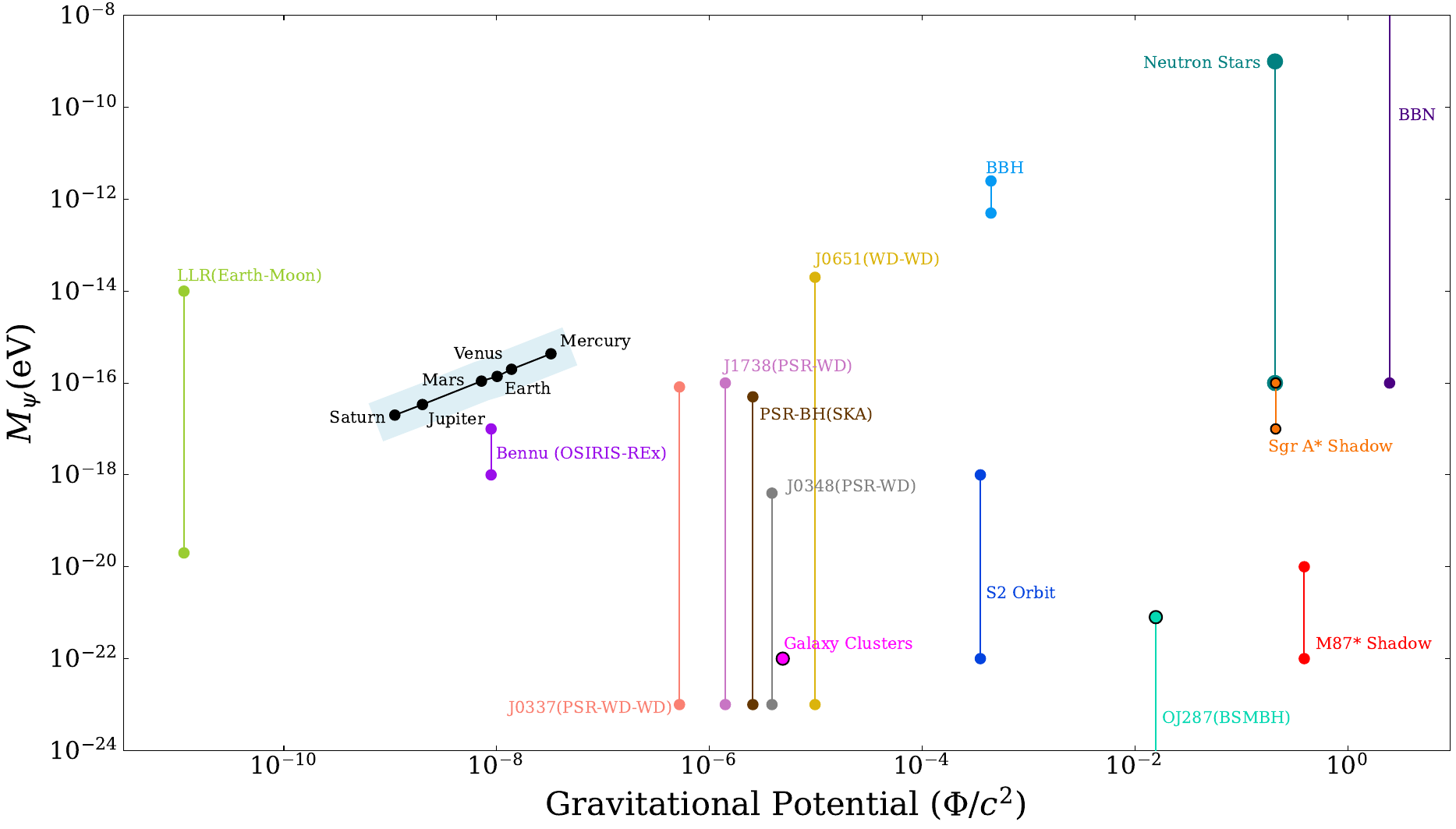}
    \caption{Bounds on scalaron mass in different astrophysical and cosmological systems with respect to gravitational potential presented by the systems. The bounds depicted in the figure include constraints on scalar field mass arising from shadow measurements of GC black hole (Sgr A*) \citep{kalita2023constraining,Paul_2024}, M87 black hole (M87*) \citep{Hui_2019}, S2 star orbit around GC black hole \citep{doi:10.1142/S0218271823500219,10.1093/mnras/stz2300}, binary pulsar-black hole studies of Square Kilometer Array (PSR-BH(SKA)) \citep{10.1093/mnras/stu1913}, pulsar-white dwarf binaries(PSR-WD) J1738 \citep{10.1111/j.1365-2966.2012.21253.x}, J0348 \citep{doi:10.1126/science.1233232}, white dwarf-white dwarf binary (WD-WD) J0651 \citep{Hermes_2012}, stellar triple system J0337 (PSR-WD-WD) \citep{Seymour_2020}, neutron stars \citep{PhysRevD.93.084038,PhysRevD.110.084011}, binary black holes (BBH) \citep{PhysRevLett.133.121401},test of inverse square law through Lunar Laser Ranging (LLR) experiment \citep{PhysRevLett.93.261101,PhysRevD.102.104003}, the OSIRIS-REx mission and ground-based tracking data for the asteroid Bennu \citep{Tsai:2023zza}, galaxy clusters \citep{10.1093/mnras/stx651}, Big Bang Nucleosynthesis (BBN) \citep{Talukdar_2024} and binary supermassive black holes (BSMBH) in the quasar system OJ287 \citep{PhysRevD.100.024010}. The black line inside the shaded area on the left represents the mass of scalarons obtained in this work. }
    \label{fig6}
\end{figure}

Constraint on scalar field mass is an interesting avenue for observational test of modified gravity theories and dark matter candidates. There are several astrophysical and cosmological probes for the scalar mass \citep{10.1093/mnras/stx651,PhysRevD.102.104003}. Studies of spontaneous scalarization of neutron stars in presence of massive scalar field have produced the bound ($10^{-16}-10^{-9}$) eV \citep{PhysRevD.93.084038,PhysRevD.110.084011}. Studies of pulsar-black hole (PSR-BH) binaries to be carried out by Square Kilometer Array (SKA) have given the bound ($10^{-23}-10^{-16}$) eV \citep{10.1093/mnras/stu1913,PhysRevD.102.104003}. Several constraints from pulsar-white dwarf (PSR-WD) \citep{10.1111/j.1365-2966.2012.21253.x,doi:10.1126/science.1233232}, white dwarf - white dwarf (WD-WD) \citep{Hermes_2012} binary systems and test of strong equivalence principle (SEP) in triple star system (pulsar-white dwarf-white dwarf) \citep{Seymour_2020} also exist. In larger scales of cosmology scalar field replaces cold dark mater (CDM) and is found to be compatible with small scale dark matter mass distribution of galaxy clusters for scalar mass $10^{-22}$ eV \citep{10.1093/mnras/stx651}. Consideration of Big Bang Nucleosynthesis (BBN) in presence of $f(R)$ gravity scalarons have produced the bound $(10^{-16}-10^{4})$ eV for compatibility with light element abundance \citep{Talukdar_2024}. The binary supermassive black hole system in the quasar OJ287 at redshift $z\sim 0.306$ has produced sufficiently light scalaron mass ($<10^{-22}$ eV) through the consideration of scalar charge induced on black holes surrounded by time dependent background scalar field \citep{PhysRevD.100.024010}. The constraints on scalar field mass arising from the orbit of the S2 star around the GC black hole and shadow measurements of the GC black hole (Sgr A*) and the M87 black hole (M87*) reported by the Event Horizon Telescope are also considered \citep{Hui_2019}. The mass bound for scalarons obtained in this study and the bounds on scalar field mass arising from the above mentioned as well as several other independent probes are presented in Figure \ref{fig6} with respect to magnitude of gravitational potential presented by these probes.

\section{Results and discussion}\label{sec5}

In this work, we have tested consistency of f(R) gravity scalarons in the scale of the solar system. We have used the observational bounds on perihelion shift of the solar system planets (till Saturn) to constrain mass of the scalarons for these planetary orbits. The masses thus obtained have been presented in Table \ref{tab2}. It is seen that scalaron mass is sensitive to scalaron field amplitude ($\psi_o$) for every orbit. We check the consistency of these mass bounds with the PPN parameter $\gamma$ and is presented in Table \ref{tab3}. It is found that low mass scalarons ($10^{-22}-10^{-18}$) eV cause significant deviation from general relativity ($\delta\gamma/\gamma\approx 0.49$). For two values of scalaron field amplitude ($\psi_o=1$ and $\psi_o=20.833$) it is found that high mass scalarons ($10^{-17}-10^{-16}$) eV are compatible with general relativity ($\delta\gamma/\gamma\approx 10^{-5}$). We reproduce these masses using observational bounds on Dicke parameter $\omega_{BD}$. It has been found that ($10^{-17} - 10^{-16}$) eV scalarons produce compatibility with general relativity in the solar system. Also, these scalaron masses are found to uplift the minimum bound on the Dicke parameter. The connection of $f(R)$ and Brans Dicke theory in the inner solar system signifies that scalaron mass should be high.

Additionally, the coupling constant $\alpha$ was constrained by \citet{Lorenzo_Iorio_2007} as $\alpha = 10^{-9}$. This was achieved through secular rate of change of perihelia of Mercury and Earth. This bound corresponds to scalaron field amplitude $\psi_o =3.33\times 10^8$. Despite such high value of $\psi_o$, the resulting scalaron mass ($M_\psi$) is nearly the same as the one obtained from $\psi_o = 20.833$ (See Table \ref{tab2}). Recently, data from OSIRIS-REx mission and ground based tracking of the asteroid Bennu have been used to derive constraints on mass of scalar mode mediating a fifth force as ($10^{-18}-10^{-17}$) eV. This falls near the mass range of fuzzy dark matter \citep{Tsai:2023zza}. In the scale of the asteroid ($\sim 1$ au) our bound on scalaron mass is ($10^{-17}-10^{-16}$) eV which is stringent relative to those of \citet{Tsai:2023zza} (See the discussion in section \ref{sec3}). This mass bound is also stringent relative to the bound of ultralight scalar which acts like dark matter candidate. For instance CMB anisotropy measurements call for $m\geq 10^{-24}$ eV \citep{PhysRevD.91.103512} which is tightened to $m\geq 10^{-23}$ eV by inclusion of weak gravitational lensing \citep{10.1093/mnras/stac1946}.

\par The mass scale derived in this work corresponds to scale of the planets in the solar system. $10^{-16}$ eV scalarons were earlier predicted to exist near the horizon of the GC black hole which affect pericenter shift of compact stellar orbits near the black hole \citep{Kalita_2020}. In a recent study \cite{Paul_2024} reported that these scalarons show GR like behavior of the gravity theory by reproducing angular size of the bright emission ring of the GC black hole shadow. The shadow measurements near GC black hole with the masses $10^{-16}$ eV and $10^{-17}$ eV are highly compatible with GR. In the inner solar system as well these masses are found to be compatible with GR. Therefore, scalaron masses constrained in the solar system are compatible with those studied near the GC black hole.

\par We now quantitatively justify that the scalarons realised in the solar system satisfy the GR criteria $M_\psi r>>1$. $0.1$ meV scalarons carry a Compton wavelength of $0.1$ cm. This is much smaller than size of the Sun's core, the environment presenting this type of scalarons. $10^{-16}$ eV and $10^{-18}$ eV scalarons have Compton wavelengths of $10^{11}$ cm and $10^{13}$ cm respectively. These are scalarons obtained from planetary orbits ranging from Mercury's to Saturn's. The orbital sizes concerned are $10^{12}$ cm ($\sim 0.3$ au for Mercury's orbit) and $10^{14}$ cm ($\sim 9.5$ au for Saturn's orbit). Therefore these scalarons are compatible with GR in the solar system.

\par Scalarons realized from solar system consideration are projected in cosmological context. Assuming the universe as a black hole having an epoch independent equality between its gravitational radius and Hubble length and using a novel relation between scalaron mass and black hole mass we generate a time varying scalaron mass. These cosmological scalarons meet those arising from solar system consideration 
($9.29\times 10^{-18} - 5.64\times 10^{-16}$) eV in the radiation era ($t = 0.88$ sec $-\ 53.89$ sec after the Big Bang). 

\par We compare scalaron masses constrained in this work with scalar field mass obtained by various astrophysical and cosmological probes. Variation of scalar mass bound with the environment of gravitational potential of different probes has been studied and the scalaron mass obtained in this study is superimposed. The $M_\psi$ range ($10^{-18}-10^{-16}$) eV obtained in this work is compatible with the ranges obtained through the study of binary pulsar systems, orbit of the S2 star near the GC black hole and the shadow measurements of GC black hole and M87 black hole (see Figure \ref{fig6}). The bounds obtained also fall within the bound ($10^{-23}-10^{-14}$) eV given by  LIGO-Virgo Catalog GWTC-1 of binary black holes \citep{PhysRevD.102.104003,PhysRevD.100.104036}. The Yukawa scale for the scalaron mass obtained in the study is in the range ($0.08-8.25$) au. The most massive scalaron ($10^{-16}$ eV), therefore, is hidden deep inside the orbit of Mercury ($\sim 0.3$ au). The scalaron mass $10^{-18}$ eV corresponds to a scale which is very close to the size of Saturn's orbit. This is compatible with the hypothesis that the range of Yukawa fifth force touches orbital scale in the solar system.

\section{Conclusion}\label{sec8}

We conclude with the following lines. $f(R)$ gravity theory with scalarons is found to be consistent in the solar system. In the solar system the theory reduces to GR with scalaron mass ($10^{-18}-10^{-16}$) eV which satisfy GR criterion $M_\psi r>>1$. From environmental dependence of scalaron mass, allowed by perturbation of a novel relation between scalaron mass and Hubble parameter we demonstrate that these are perturbed scalarons in the environment concerned. We reproduce them as background scalarons in the radiation era of the universe and infer that scalaron mass evolves with cosmic time. The window $(10^{-18}-10^{-16})$ eV is found to be compatible with existing bounds on scalar field mass produced by several astrophysical and cosmological probes. It has been possible to extract a mass range of the $f(R)$ gravity degree of freedom which is compatible with the mass range derived from the effect of the theory at the neighborhood of the GC black hole and binary pulsar systems. $f(R)$ gravity is a serious alternative to GR. More precise measurements in the outer solar system regime in future is expected to better constrain the theory.

\section*{Acknowledgments}
The authors would like to acknowledge the anonymous reviewers for prompt and helpful suggestions.

\bibliography{reference}{}
\bibliographystyle{mnras}

\end{document}